\documentclass{article}

\usepackage{graphicx}
\usepackage{url}
\usepackage{tabularx}
\usepackage{sidecap}
\usepackage{rotating}

\begin{document}

\title{What are essential concepts about networks?}

\author{
Hiroki Sayama$^*$\\
Center for Collective Dynamics of Complex Systems,\\
Binghamton University, State University of New York, Binghamton, NY 13902-6000, USA\\
Center for Complex Network Research, Northeastern University, Boston, MA 02115, USA\\
$^*$Corresponding author: sayama@binghamton.edu\\
\and
Catherine Cramer\\
New York Hall of Science, Corona, NY 11368, USA\\
\and
Mason A. Porter\\
Oxford Centre for Industrial and Applied Mathematics,\\
Mathematical Institute, University of Oxford, OX2 6GG, UK\\
CABDyN Complexity Centre, University of Oxford, Oxford, OX1 1HP, UK\\
\and
Lori Sheetz\\
Network Science Center, United States Military Academy at West Point,\\
West Point, NY 10996, USA\\
\and
Stephen Uzzo\\
New York Hall of Science, Corona, NY 11368, USA
}

\maketitle

\begin{abstract}
Networks have become increasingly relevant to everyday life as human society has become increasingly connected. Attaining a basic understanding of networks has thus become a necessary form of literacy for people (and for youths in particular). 
 At the NetSci 2014 conference, we initiated a year-long process to
  develop an educational resource that concisely summarizes essential
  concepts about networks that can be used by anyone of school age or older. 
  The process involved several brainstorming sessions on one key question: {\em
    ``What should every person living in the 21st century know about
    networks by the time he/she finishes secondary education?''} 
    Different sessions reached diverse participants, which included professional researchers in network science, educators, and high-school students. The generated ideas were
  connected by the students to construct a concept
  network.  We examined community structure in the concept network to group ideas into a set of important themes, which we refined through discussion into seven essential concepts. The students played a major role in this development process by providing insights and perspectives that were often unrecognized by researchers and
  educators. The final result, {\em ``Network Literacy: Essential
    Concepts and Core Ideas''}, is now available as a booklet
  in several different languages from
  \url{http://tinyurl.com/networkliteracy}.
\end{abstract}

\section{Introduction}\label{introduction}

Network science has matured over the past few decades, and its potential importance for improving understanding of complex natural and human-made phenomena is now recognized in an increasingly diverse set of domains \cite{newman2010,barabasi2014linked}. 
Its potential beneficiaries include not only researchers, but also business
practitioners, policymakers, educators, and virtually all lay people, as they need to understand and deal with complex real-world phenomena. To advance the use of networks as a lens on ourselves and on our world,
there needs to be widespread and sophisticated knowledge about how the study of networks---the science of connectivity---can be an important epistemological tool for all people. In other words, we assert that an understanding of networks is a new kind of literacy that is important for everyone living in the 21st century.

We favor the term ``Network Literacy'' over ``Network Science Literacy'', because we believe that the scope and impact of networks goes beyond what one would typically
construe as ``science''. The scope of Network Literacy is also rather different from what is currently being taught as ``Digital Literacy'' or ``Internet Literacy'' about the use of computers, the Internet, and other digital
media \cite{digital,internet}. Network Literacy needs to include fundamental concepts about
networks, including (1) how an overwhelming abundance of everyday phenomena can be viewed
through the lens of connectivity, interactions, and interdependence; and
(2) how one can analyze, understand, utilize, and improve their features.

In response to such emerging societal needs, researchers in the
network-science community have started educational outreach efforts to bring network science to
secondary education
\cite{harrington2013commentary,sanchez2014more,cramer2015netsci}. 
There are also a few other dispersed outreach efforts that have considered, or even focused on, teaching ideas about networks (see, e.g., \cite{budd,meeks,nets-outreach1,tangiblenetworks,virustracker}).
However, there do not yet exist educational materials about networks that are both systematically structured and easily
accessible---such as grade-school-level textbooks, workbooks, curricular modules, and lesson plans
(e.g., the educational materials \cite{quaden2005shape,potash2011dollars,anderson2014systems} developed by the ``System Dynamics'' community \cite{forrester1997industrial,sterman2000business})---and which
can be integrated readily into formal and informal educational
programs. As a first step towards meeting this need, we
undertook a year-long effort to develop a set of Essential Concepts for Network Literacy.

We started by consulting people who had succeeded in similar
educational initiatives. We were particularly inspired by the ``Ocean
Literacy'' initiative
\cite{cava2005science,schoedinger2006need,strang2007can}, which was spearheaded by members of
the US NSF Centers for Ocean Science Education Excellence (COSEE). The
work on the Ocean Literacy initiative began in 2002 when a grassroots
group of scientists and educators, frustrated by the lack of
ocean-science content in national education standards, came together
to discuss what they felt every person should know about the ocean by the time
he/she graduates from high school.\footnote{Their history is discussed at \url{http://oceanliteracy.wp2.coexploration.org/ocean-literacy-network/foundations/history/}.} After discussing initial ideas for two years, the group organized a two-week online conference in 2004 that included more than 100 participants, a live
keynote talk and virtual ``rooms'' organized according to scientific disciplines and age groups. Their next step was to hold a charrette, at which 30 people convened for two days and distilled the combined input into 7
principles (plus 44 fundamental concepts to more fully explain these
principles). The final result was published in 2005 as a pamphlet called
``Ocean Literacy: The Essential Principles and Fundamental Concepts of
Ocean Sciences for Learners of All Ages''.\footnote{See \url{http://www.oceanliteracy.net/}.} To date, more than 30,000 brochures have
been distributed in hard copy (in addition to digital dissemination). The success of Ocean Literacy was
followed by similar processes in Climate Literacy, Atmospheric Literacy, Earth Science Literacy, Great Lakes Literacy, Energy Literacy, and more.\footnote{See \url{http://nagt.org/nagt/teaching_resources/literacies.html}.}

Our goal is similar: we wish to provide a simple, concise, high-level guiding document that can facilitate the process of developing resources for disseminating Network Literacy to all people. To initiate this
process, we posed one key question to ourselves and to the network-science community: {\em What should every person living in the 21st century know about networks by the time he/she finishes secondary education?}

In the present paper, we describe the highly collaborative process by which we developed the first
version of ``Network Literacy: Essential Concepts and Core Ideas'' (which is now available at \url{http://tinyurl.com/networkliteracy}). Network-science researchers, educators, high-school students, and others collectively explored a wide variety of ideas and developed a network of concepts.
We examined community structure \cite{porter2009} in the network of concepts to obtain several concept clusters, which were further discussed and refined to ultimately yield seven essential concepts about networks. We hope that the process and outcomes of this
initiative will be useful for people who are interested in network science and in networks more generally, and we also hope that it will have positive effects on teaching, learning, communicating, and policy-making about networks.

\section{Methods}\label{methods}

\subsection{Brainstorming and Concept Generation}

\begin{figure}[p]
\centering
\includegraphics[height=0.95\textheight]{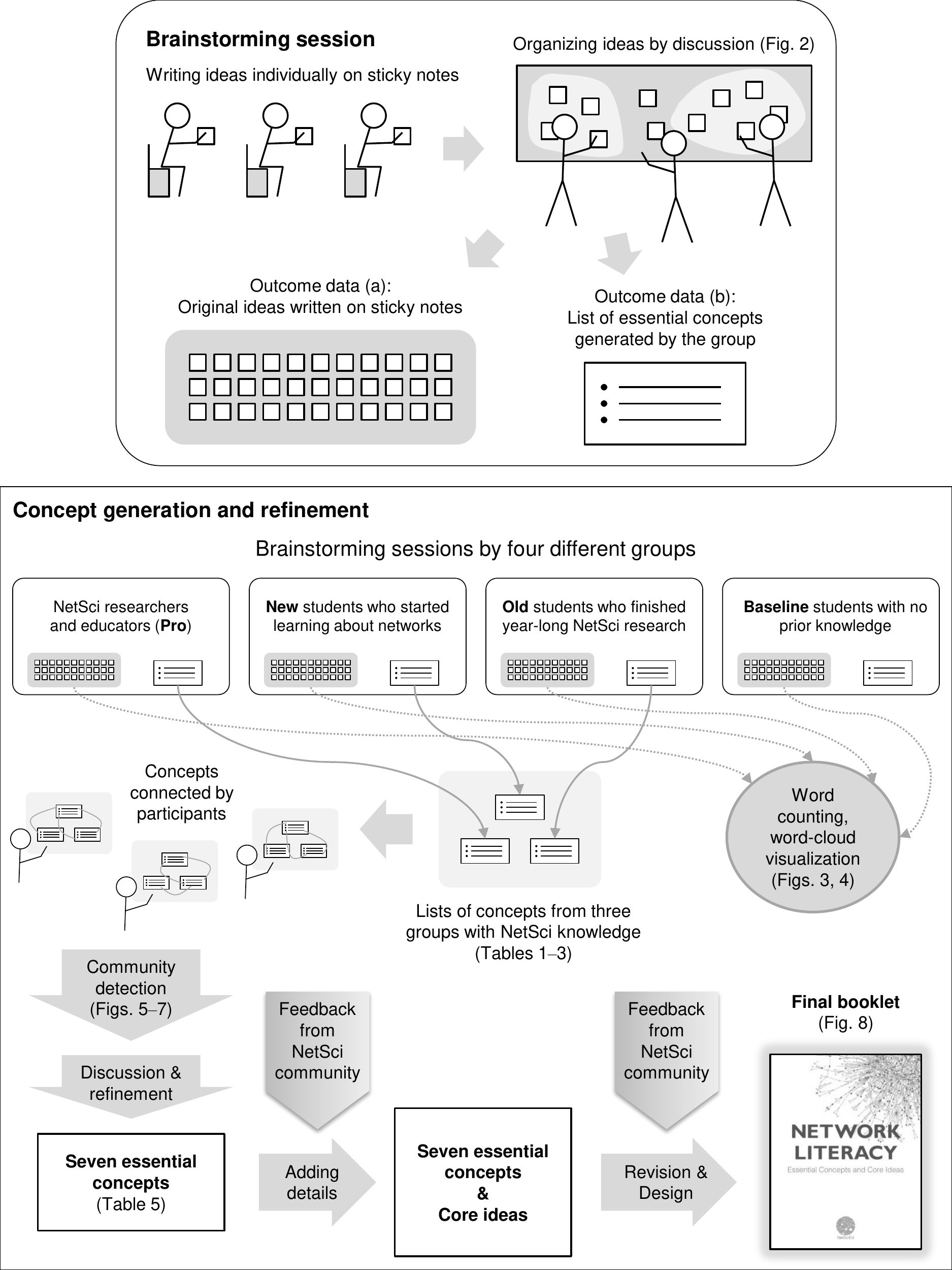}
\caption{Overview of the work flow of the process by which the
  ``Network Literacy'' booklet was developed. (Top) Procedure of each
  initial brainstorming session. (Bottom) Process of concept generation
  and refinement.
  }
\label{fig:workflow}
\end{figure}

\begin{figure}[tbp]
\centering
\includegraphics[width=0.9\textwidth]{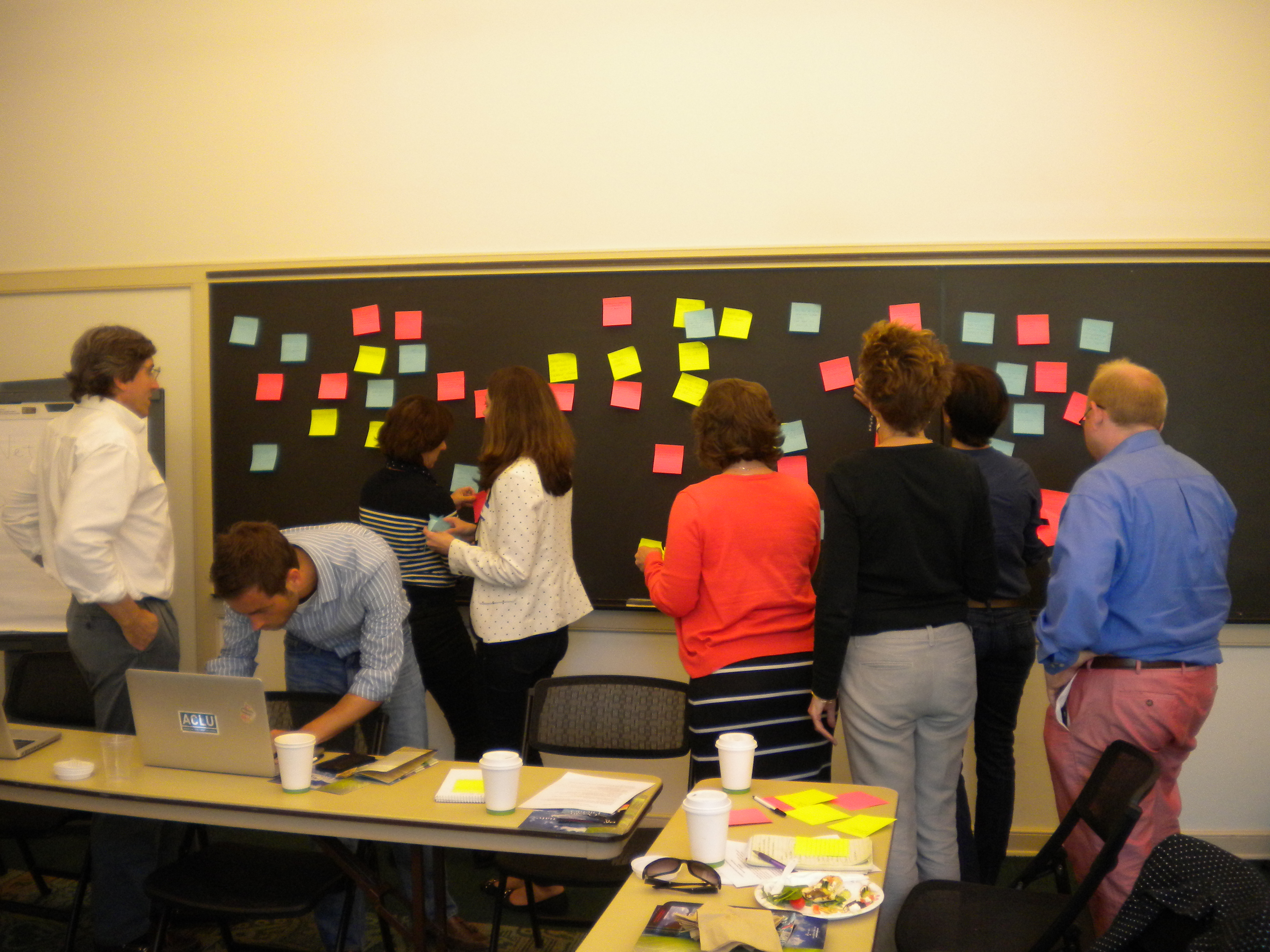}
\caption{Initial brainstorming session by network-science researchers
  and educators at the NetSci 2014 pre-conference event held at
  University of California at Berkeley on 1 June 2014. No high-school students were present at this session. Participants
  posted their ideas of essential concepts about networks using sticky
  notes, and they then organized them into several concept clusters
  through discussion. The same format was used for the other brainstorming sessions
  with high-school students. (Photo credit: Hiroki Sayama)}
\label{fig:brainstorming}
\end{figure}

Our starting point was to hold several brainstorming sessions to explore various
concepts and ideas about networks that could be used for developing
the list of essential concepts (see the top panel of Fig.~\ref{fig:workflow}). Each session followed the same basic procedure:
\begin{enumerate}
\item Have each participant introduce him/herself to the rest of the group.
\item Explain the objective of the session and then pose the
  following key question to the group: ``What should
  every person living in the 21st century know about networks by the
  time he/she finishes secondary education?'' 
\item Hand out sticky notes to the participants.
\item Let each participant write his/her ideas about what ``network
  literacy'' might mean on the sticky notes. Participants work on this phase as
  individuals without communication. This usually takes about 15 minutes.
\item Have participants place their sticky notes on a large wall or a
  blackboard, and then have them cluster the notes into several topical areas
  based on their similarity and relatedness. This phase usually takes 15 minutes or often longer, and it involves a lot
  of physical interaction and conversations
  among the participants (see Fig.~\ref{fig:brainstorming}). 
\item Have a group discussion on the result of brainstorming to
  refine and distill a final set of essential concepts. This phase
  also takes about 15 minutes.
\item Break the group into subgroups and have each subgroup
  visualize one of the essential concepts that arise from the
  brainstorming. For educational purposes, we included this phase only at sessions with high-school
  students. 
  Their visualization results were later compiled into a PowerPoint presentation and
  presented to the students to summarize and reinforce
  their learning experience.
 \end{enumerate}

Each brainstorming session produced two primary sets of outcome data:
(a) handwritten data on sticky notes; and (b) a final compiled list of
essential concepts that the group produced (inspired by the sticky notes) through brainstorming. We used the
former to qualitatively compare the outputs of different participant groups,
and we used the latter to develop a final version of the essential
concept list.

We ran brainstorming sessions with the following four participant groups (see the bottom panel of Fig.~\ref{fig:workflow}):
\begin{itemize}
\item Network-science researchers and educators who participated in
  NetSci 2014 (``professionals'' or ``pro''). The session was run at
  a NetSci 2014 pre-conference event at University of California at
  Berkeley on 1 June 2014. Unlike the other three sessions, this group also had a
  follow-up discussion during the next day at the NetSciEd3 satellite symposium\footnote{See \url{https://sites.google.com/a/nyscience.org/netscied3/}.} to revise the list of essential
  concepts (and to get input from people who could not attend the previous day).
\item The 2014--2015 cohort of NetSci High \cite{cramer2015netsci} high-school students who just finished an initial training workshop on network science (``new students'' or ``new''). The session was run
  at the NetSci High Summer Workshop at Boston University on 22 July 2014.
\item The 2013--2014 cohort of NetSci High high-school students who
  had completed a year-long research experience in network science (``old
  students'' or ``old''). The session was run at the NetSci High
  Summer Workshop at Boston University on 23 July 2014.
\item Explainers at the New York Hall of Science (mostly high-school
  students, but also some university students) who did not have any learning or training
  experience about networks (``baseline group''). The session was run
  at the New York Hall of Science on 13 August 2014.
\end{itemize}

\subsection{Comparison Between Participant Groups}

To compile raw text data from each group, the original ideas written on the sticky notes by the participants
were collected and typed up. We conducted simple content analysis of this data set based on
word-frequency counting
\cite{weber1990basic,stemler2001overview}. Specifically, we broke the text data into words, and we then removed trivial words (e.g.,
articles and simple prepositions). We converted the remaining
words into a canonical form by converting plural
nouns to singular ones and by removing aspects and tenses from
verbs. We also removed words that appeared only once in the data for each group.

For initial visual inspection, we used Mathematica 10.2's {\tt WordCloud} function (in its default settings) to construct a word cloud for the words generated by each group. We then compared the word frequencies across the four
groups to detect semantic differences between them. We generated word rankings based on their frequencies and visualized them in Mathematica 10.2 using a custom script. Because of the small number of samples, we did not conduct statistical tests.

\subsection{Development of Final List of Essential Concepts}

We used the lists of essential concepts produced by the three participant
groups with some experience in network science (``professionals'',
``old students'', and ``new students'') to help develop a final
list of essential concepts about networks (see the bottom panel of Fig.~\ref{fig:workflow}). To facilitate this process,
we presented the participants (researchers, educators, and high-school students) at
the NetSci High Summer Workshop with a sheet of paper on which {\em all} of the concepts from the
three groups were presented together, and we asked them to draw connections among
those concepts.

From the results of this activity, we produced a single multigraph, in which each essential concept produced by the three groups is represented by a node and in which an edge represents a connection that a participant made between a pair of concepts. 
We detected communities of concepts on this multigraph using Mathematica 10.2's {\tt FindGraphCommunities} function in its default settings (which uses a method based on maximizing the modularity objective function). This produced several distinct communities with clear topical themes, as well as a patchwork community (i.e., a community whose topical theme was not clear) that consisted of the remaining concepts. 
We applied the same community-detection technique to the patchwork
community (extracted as an induced subgraph) repeatedly to attempt to discern more subtle concept clusters within it. This procedure was repeated twice (i.e., considering patchwork communities as networks and detecting communities in them), after which we were able to obtain a set of concept clusters that were thematically meaningful.

Once we detected several meaningful concept clusters, we further discussed and refined them. We then used the resulting clusters to draft seven essential concepts for Network Literacy, and we announced this list (which we posted on Google Drive) to the network-science community via e-mail to collect feedback. We incorporated the community's feedback and used it to help develop in-depth descriptions of each essential concept (called ``core ideas'') to provide more details about each concept (including
examples and implications). We then announced this version, now furnished with both essential
concepts and core ideas, to the network-science
community to receive additional feedback.  We incorporated this feedback, which we also discussed and refined further among ourselves, and placed the final content into a booklet for public distribution.

\section{Results}\label{results}

\subsection{Comparison Between Participant Groups}

In Tables \ref{tab:pro}--\ref{tab:con}, we show the lists of essential
concepts that were created by the four participant groups. There is a large difference between the
concepts created by the baseline group (i.e., high-school or university students
without any formal training with networks), whose concepts we show in Table \ref{tab:con},
and the other groups. The concepts generated by the baseline group tended to be vague and general, and they were likely derived from everyday uses of the word ``network'' in English (e.g., social networking and computer
networks). In contrast, the concepts that were created by the other three groups (see Tables
\ref{tab:pro}--\ref{tab:old}), who had learning and training experiences about networks, included much more concrete notions
about networks that likely also encapsulated a better understanding of salient ideas.
This demonstrates emphatically that education can and does have an impact on people's
understanding of networks.

There are also some interesting differences between the concepts
that were created by professionals and those that were created by students. The former
(see Table \ref{tab:pro}) generally used abstract
language and also had a clear focus on what could be done using
network models (which they also described in abstract ways). In
contrast, the concepts that were created by students (see Tables \ref{tab:new} and
\ref{tab:old}) consisted predominantly of definitions of networks and
concrete examples, together with research methodologies that they had
learned (e.g., visualization and use of computer software). These
differences convey a clear gap between professionals and lay people regarding what is perceived as most essential to the study
of networks, and they suggest that it is crucial to actively involve non-professionals (in addition to practicing network scientists) in the development of a guiding booklet of essential concepts about networks.

The Powerpoint presentations of concept visualizations that the NetSci High students constructed (see our description above) are available online.\footnote{\noindent Presentation by old students:\\ {\scriptsize
    \url{http://www.bu.edu/networks/files/2012/08/Networks-by-NetSciHigh2013-2014-student-team.pptx}}\\ Presentation
  by new students:\\ {\scriptsize
    \url{http://www.bu.edu/networks/files/2012/08/Networks-by-NetSciHigh2014-2015-student-team.pptx}}}

\begin{table}[tbp]
\centering
\caption{List of essential concepts that were created by network-science
  researchers and educators who participated in the NetSci 2014
  pre-conference event and the NetSciEd3 satellite symposium
  (``professionals'').
  }
\begin{tabularx}{\textwidth}{lX}
\hline
{\em No.} & {\em Concept}\\
1. & Networks describe how things connect \& interact.\\
2. & The world around us can be represented as networks of interconnecting parts.\\
3. & Representing networks allows them to be used as a tool.\\
4. & Modeling systems as networks can help reveal and explain patterns and general principles.\\
5. & Network structure can influence behavior and vice versa.\\
6. & Networks help us understand similarities among systems found in our everyday life.\\
7. & Networks can be represented and studied in many different ways.\\
\hline
\end{tabularx}
\label{tab:pro}
\end{table}

\begin{table}[tbp]
\centering
\caption{List of essential concepts that were created by the 2014--2015 cohort of
  NetSci High high-school students (``new students'') who just
  finished a short training workshop at the 2014 NetSci High Summer
  Workshop at Boston University.}
\begin{tabularx}{\textwidth}{lX}
\hline
{\em No.} & {\em Concept}\\
1. & Networks are connections and interactions.\\
2. & Networks are present in every aspect of life.\\
3. & Examples include economics/social/political sciences.\\
4. & Networks consist of nodes and links.\\
5. & Computers are often used to study networks.\\
6. & Networks can be used for making predictions.\\
7. & Visualization of networks helps understanding.\\
8. & Networks allows us to analyze human interactions.\\
9. & Network science should be incorporated into schools.\\
10. & Future of network science can include preinstalled programs and networks in media.\\
11. & Networks connects various aspects of the world.\\
\hline
\end{tabularx}
\label{tab:new}
\end{table}

\begin{table}[tbp]
\centering
\caption{List of essential concepts that were created by the 2013--2014 cohort of
  NetSci High high-school students (``old students'') who 
  had completed a year-long research experience.}
\begin{tabularx}{\textwidth}{lX}
\hline
{\em No.} & {\em Concept}\\
1. & Networks are everywhere.\\
2. & A connection between more than one person can form a link.\\
3. & People should understand basic terminology in a network such as nodes, edges and hubs.\\
4. & Visualizations are the key to understanding the network.\\
5. & Networks can be applied to various fields, whether through entertainment, disease spread, finding quality furniture, etc.\\
6. & Life is hard... then you die.\\
\hline
\end{tabularx}
\label{tab:old}
\end{table}

\begin{table}[tbp]
\centering
\caption{List of essential concepts that were created by high-school students
  (explainers at the New York Hall of Science) who had not had any
  learning or training experience about networks (i.e., the ``baseline
  group'').}
\begin{tabularx}{\textwidth}{lX}
\hline
{\em No.} & {\em Concept}\\
1. & People and connections\\
2. & Act of network\\
3. & Technology + networking\\
\hline
\end{tabularx}
\label{tab:con}
\end{table}

\begin{figure}[tbp]
\centering
\includegraphics[scale=0.47]{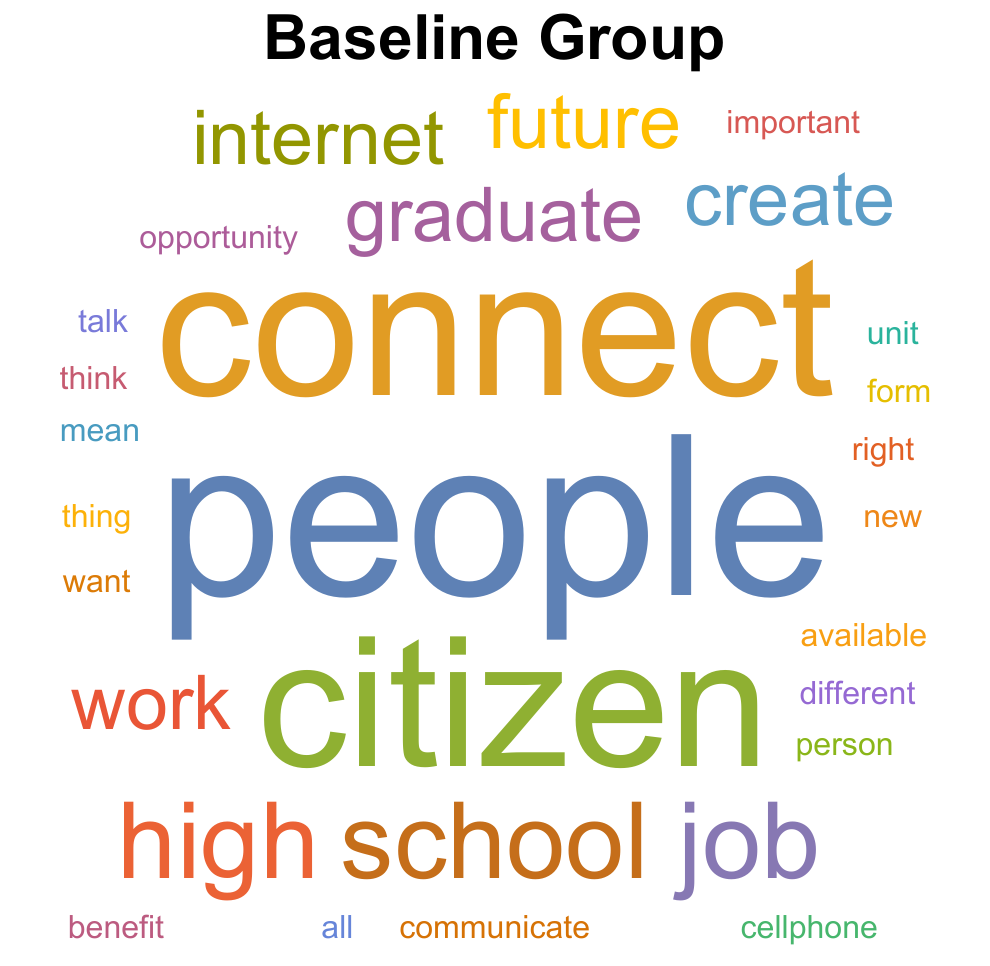}
\includegraphics[scale=0.47]{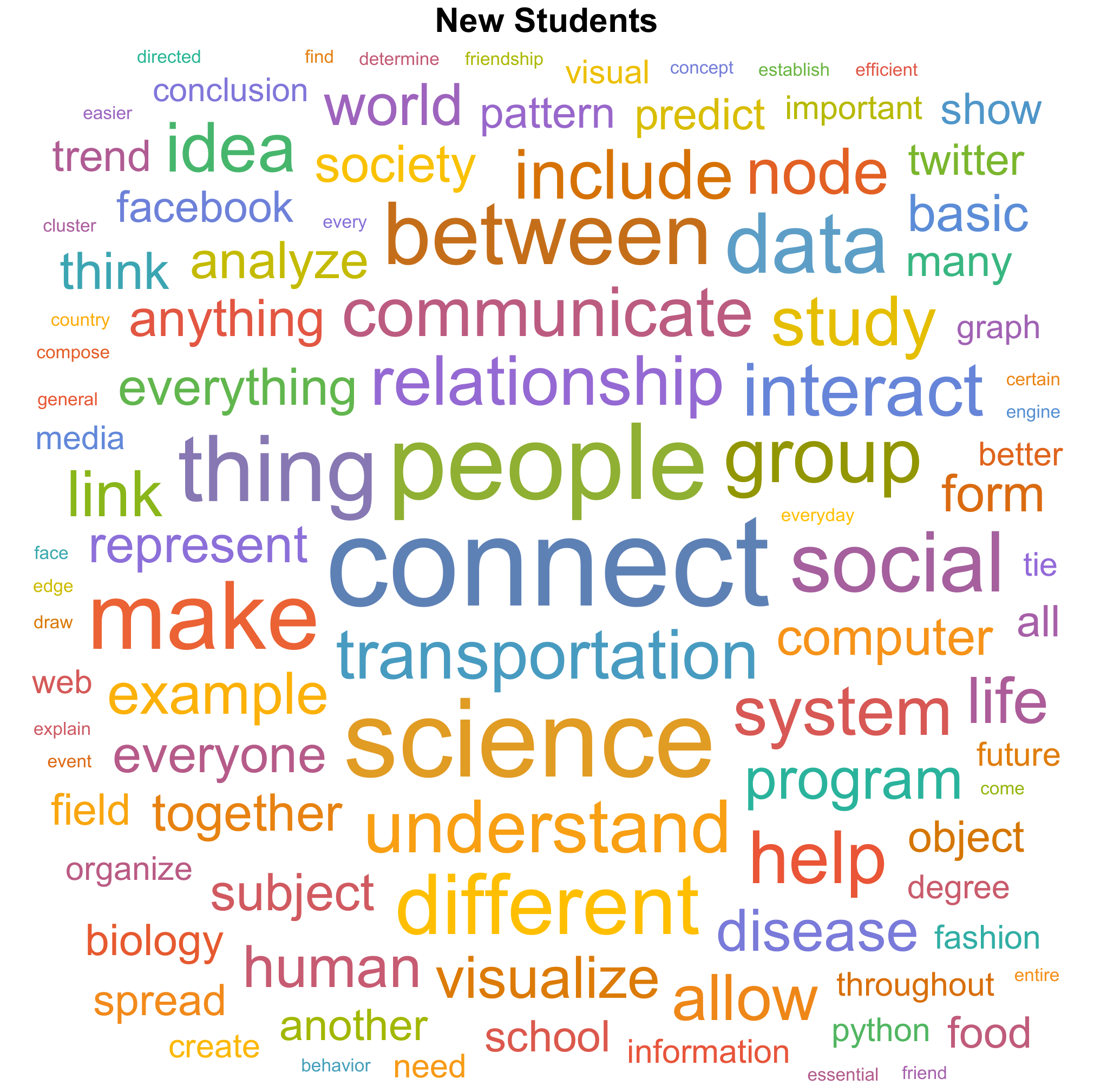}\\~\\
\includegraphics[scale=0.47]{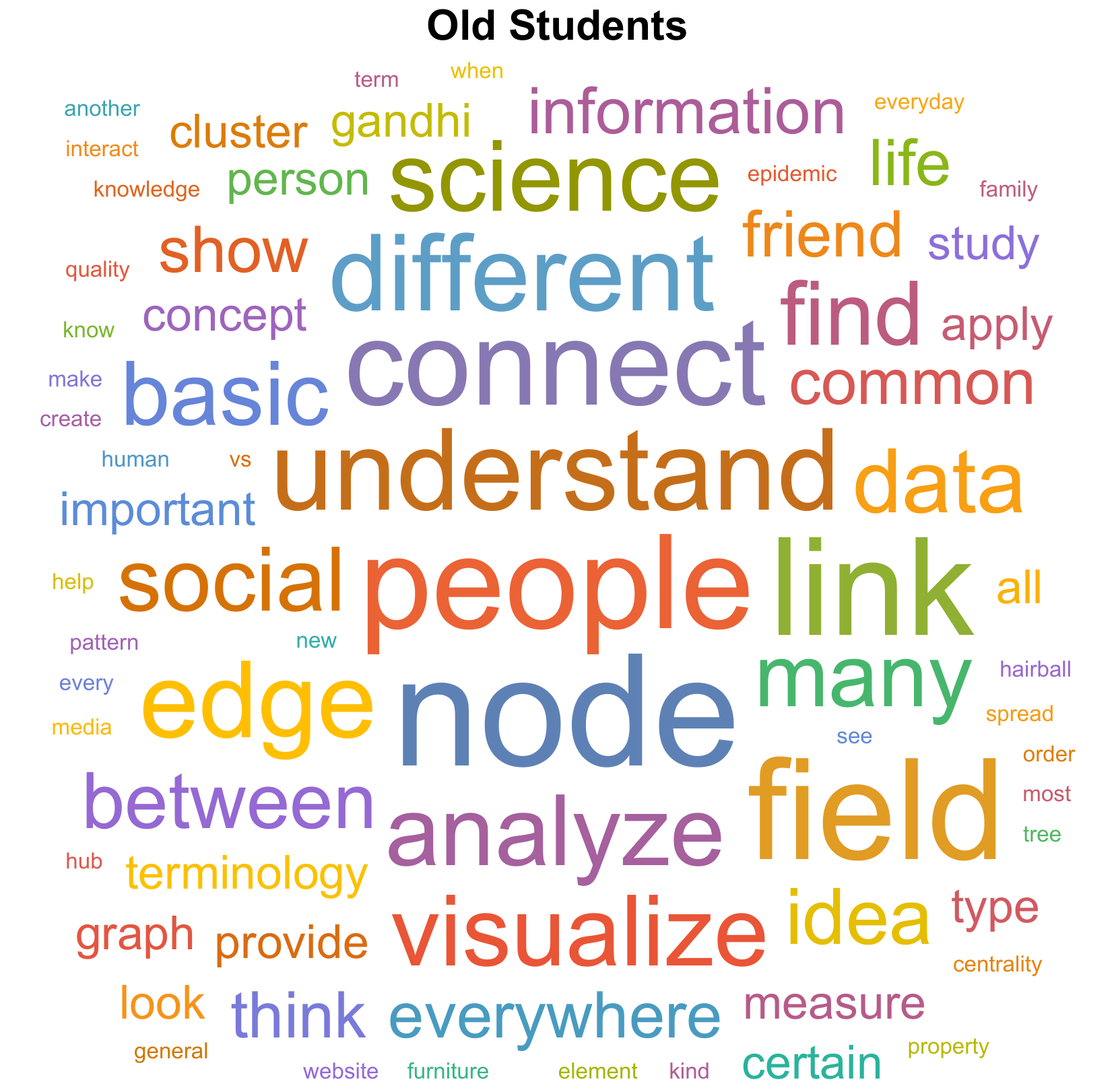}
\includegraphics[scale=0.47]{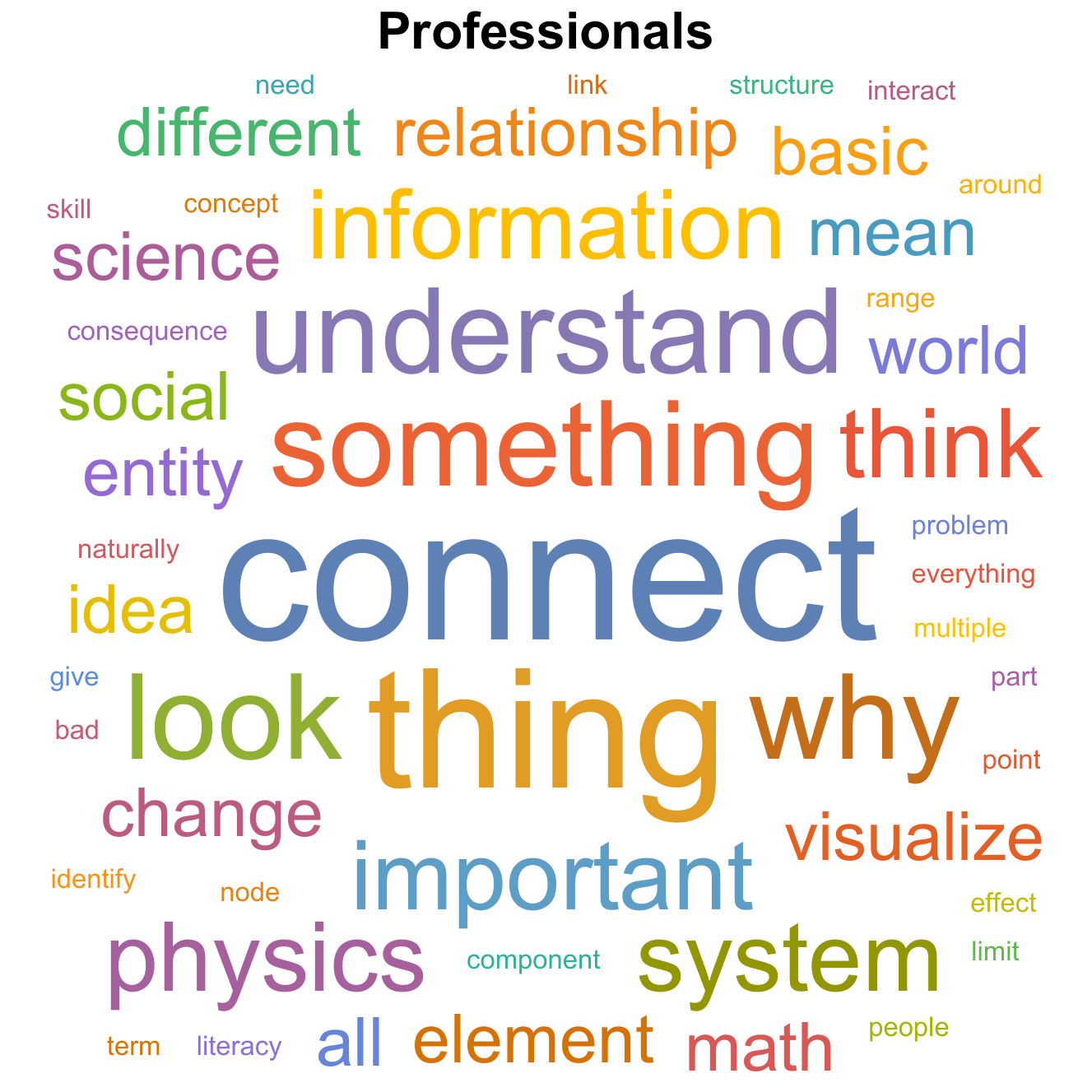}
\caption{Word clouds generated (using Mathematica 10.2) to visualize the frequencies of words used by each of the
  four participant groups (excluding ``network'', which was the most frequently-used word). We scaled the word sizes logarithmically based on their number of appearances.
  }
\label{fig:clouds}
\end{figure}

\begin{SCfigure}
\centering
\includegraphics[height=\textheight,width=0.75\textwidth]{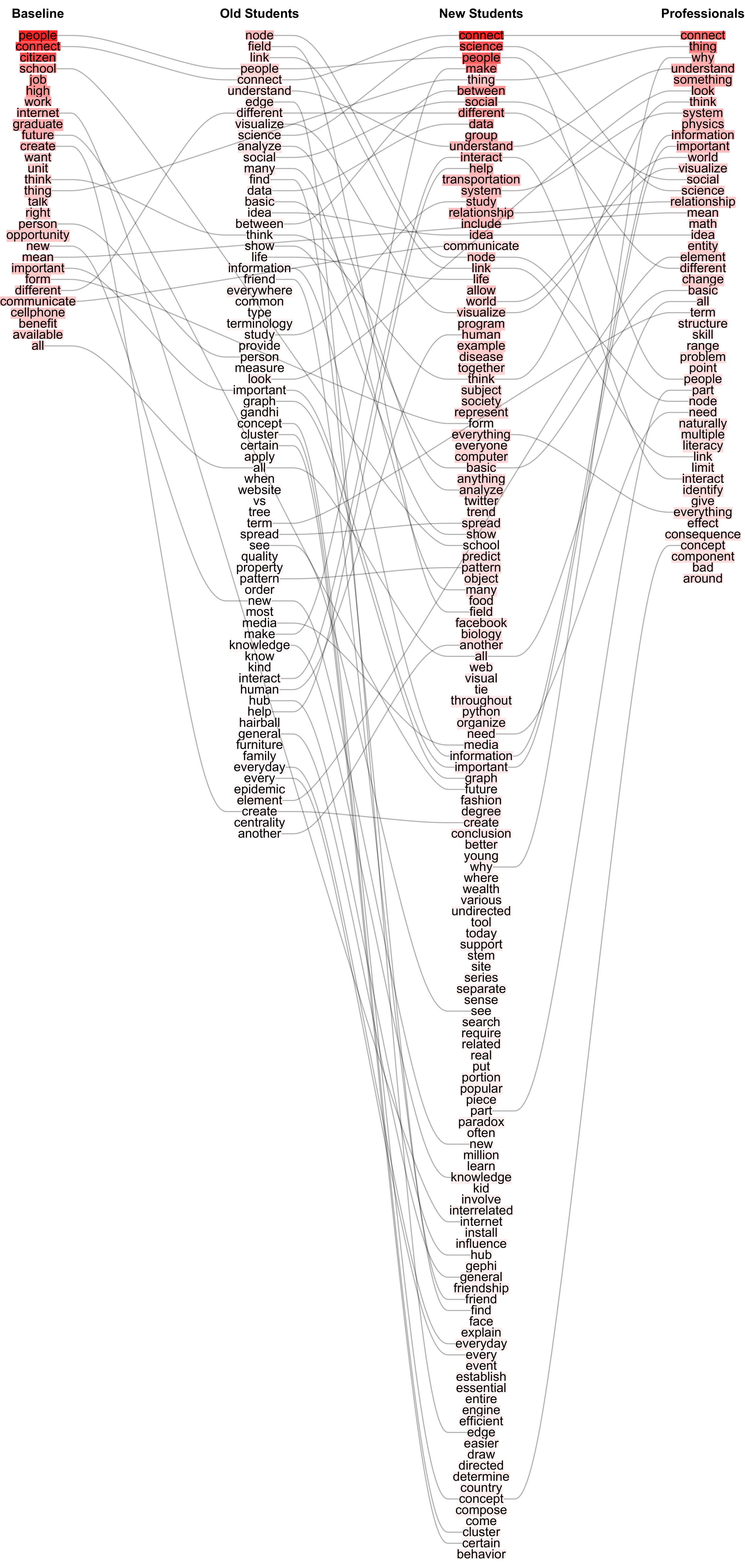}
\caption{Visualization of words from the raw text data written on sticky notes during each of the brainstorming sessions (generated by Mathematica 10.2 using a custom script). In each case, we rank the words (again excluding the word ``network'') from highest frequency at the top to lowest frequency at the bottom. The background color intensity (which is not scaled linearly)
  represents the relative frequencies of the words. When the same word appears in multiple groups, we use a curve to connect the word in the different groups.
  }
\label{fig:word-rankings}
\end{SCfigure}

Our aforementioned observations are supported further by word-cloud visualizations (see Fig.~\ref{fig:clouds}) and word-ranking
visualizations (see Fig.~\ref{fig:word-rankings}), which we 
constructed using the methods described in Section \ref{methods}.
The word ``network'' was the most frequently-used word in all four groups, and we thus removed it from the analysis. The other most common words give identifiable characterizations for each of the groups.
For example, the baseline group was the only one to include
career-related words (such as ``job'', ``work'', and ``opportunity''), which came from the participants' recognition of the importance of
``networking'' for their future career development. Additionally,
the total number of words produced by the baseline group was the
smallest, suggesting that they were not aware of the broad applicability
of network-related concepts to a wide variety of phenomena. The words
from the old students included more technical terms (e.g., ``node'',
``link'', and ``edge'') than any other group, indicating that they had
become familiar with those technical words from their year
of research experience in network science. The new students produced the
most diverse set of words (which is apparent in both word clouds and
word rankings), which nicely conveys their intellectual excitement immediately
after learning about networks. Finally, the words produced by
professionals often tended to be about high-level, abstract concepts. Interestingly, the word ``people'' was not used as often by the professionals as by other groups.

\subsection{Development of Final List of Essential Concepts}

\begin{sidewaysfigure}[p]
\centering
\includegraphics[width=\textheight]{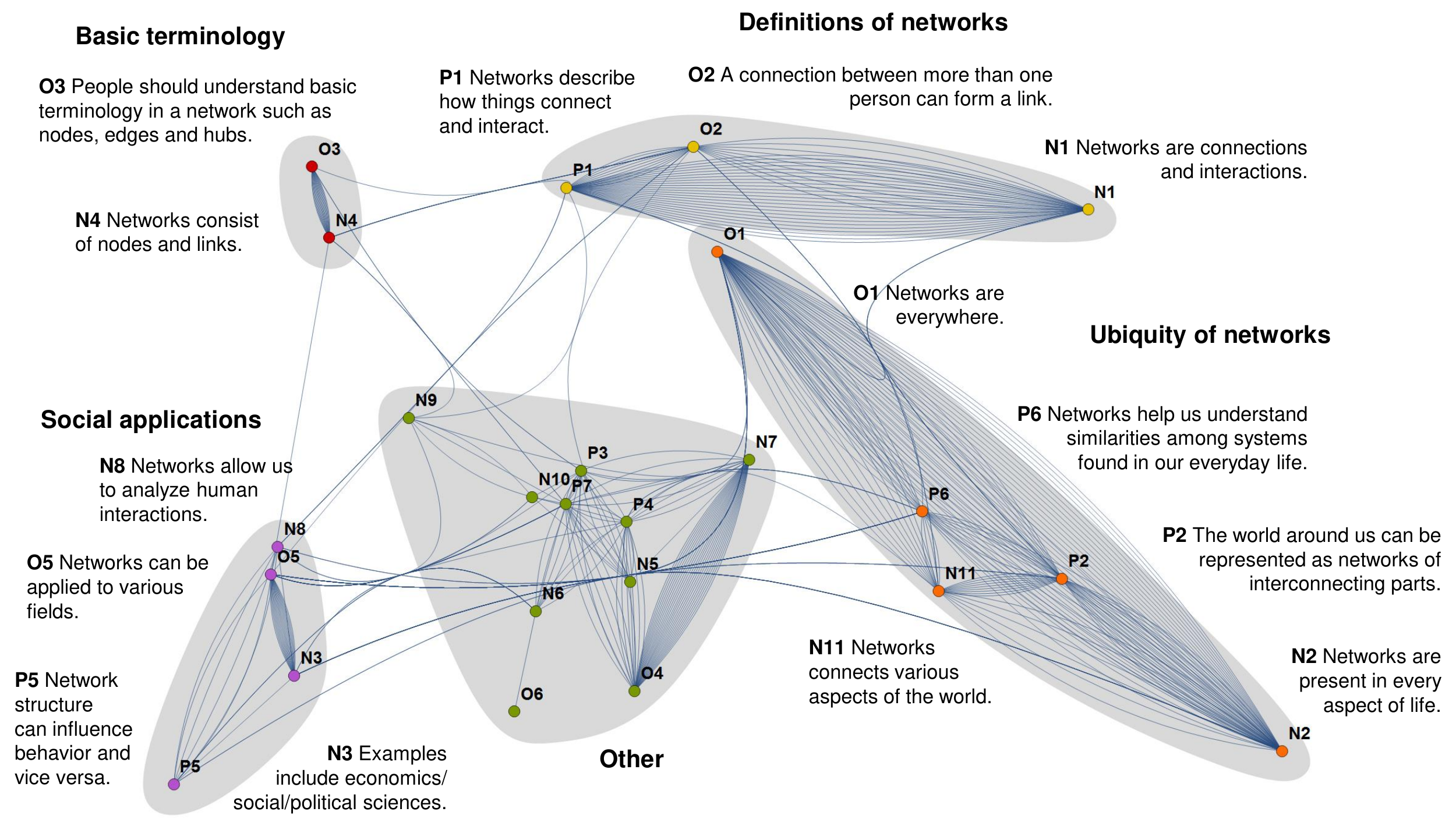}
\caption{Concept network generated by connecting concepts made by
  three groups: ``old students'' (``O''), ``new students'' (``N''), and
  ``professionals'' (``P''). The gray clouds indicate the communities that we detected using Mathematica 10.2's {\tt FindGraphCommunities} function. They represent ``Basic terminology'' (top left, old + new), ``Definitions of
  networks'' (top right, old + new + pro), ``Ubiquity of networks'' (bottom
  right, old + new + pro), ``Social applications'' (bottom left, old + new
  + pro), and ``Other''. We partition the community labeled ``Other'' in Figs.~\ref{fig:comm2} and
  \ref{fig:comm3}.
  }
\label{fig:comm1}
\end{sidewaysfigure}

\begin{sidewaysfigure}[p]
\centering
\includegraphics[width=\textheight]{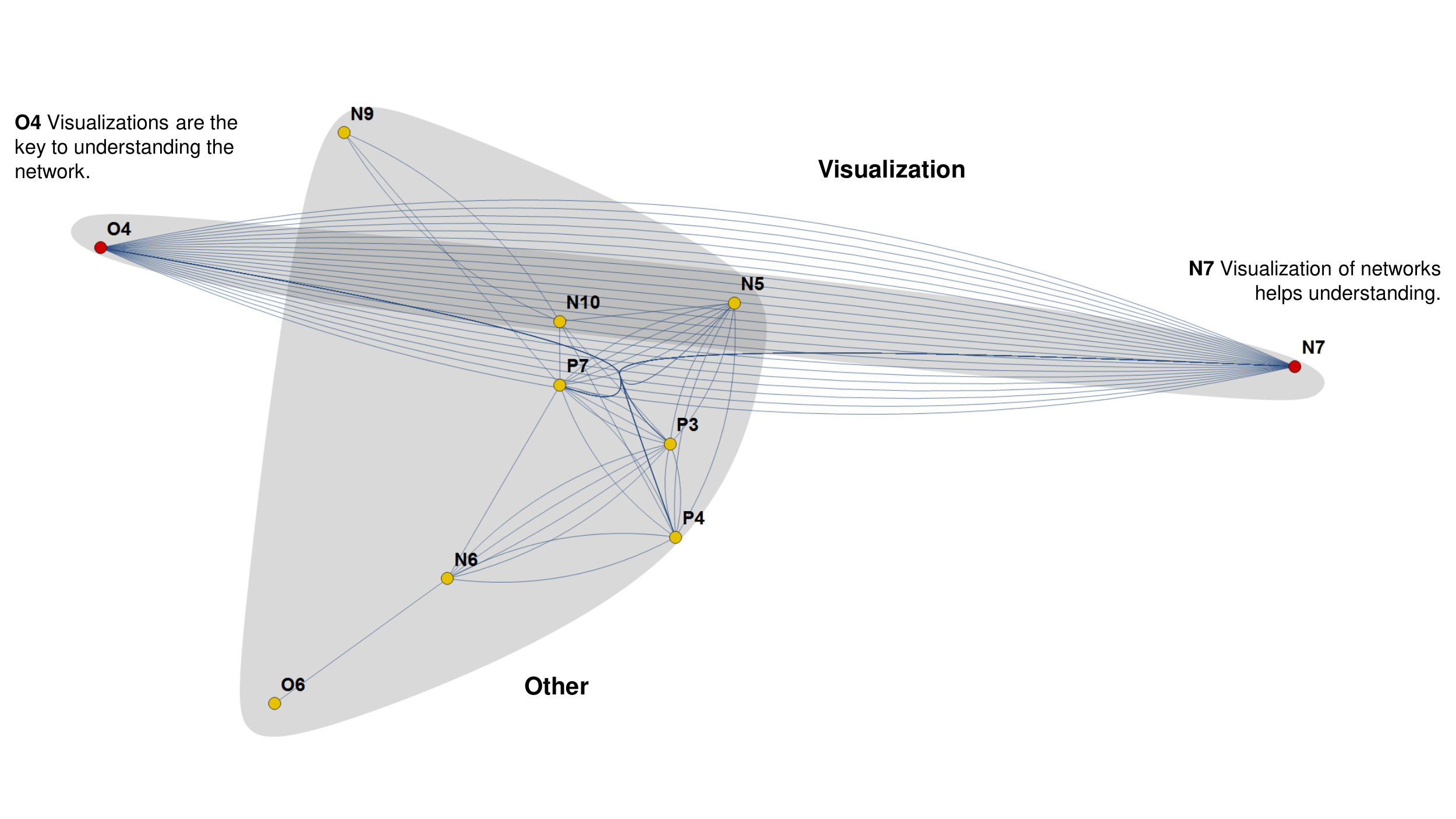}
\caption{Decomposition of the ``Other'' community in
  Fig.~\ref{fig:comm1}. We find another community about ``Visualization'' (old +
  new). We further partition the remaining ``Other'' community in
  Fig.~\ref{fig:comm3}.}
\label{fig:comm2}
\end{sidewaysfigure}

\begin{sidewaysfigure}[p]
\centering
\includegraphics[width=\textheight]{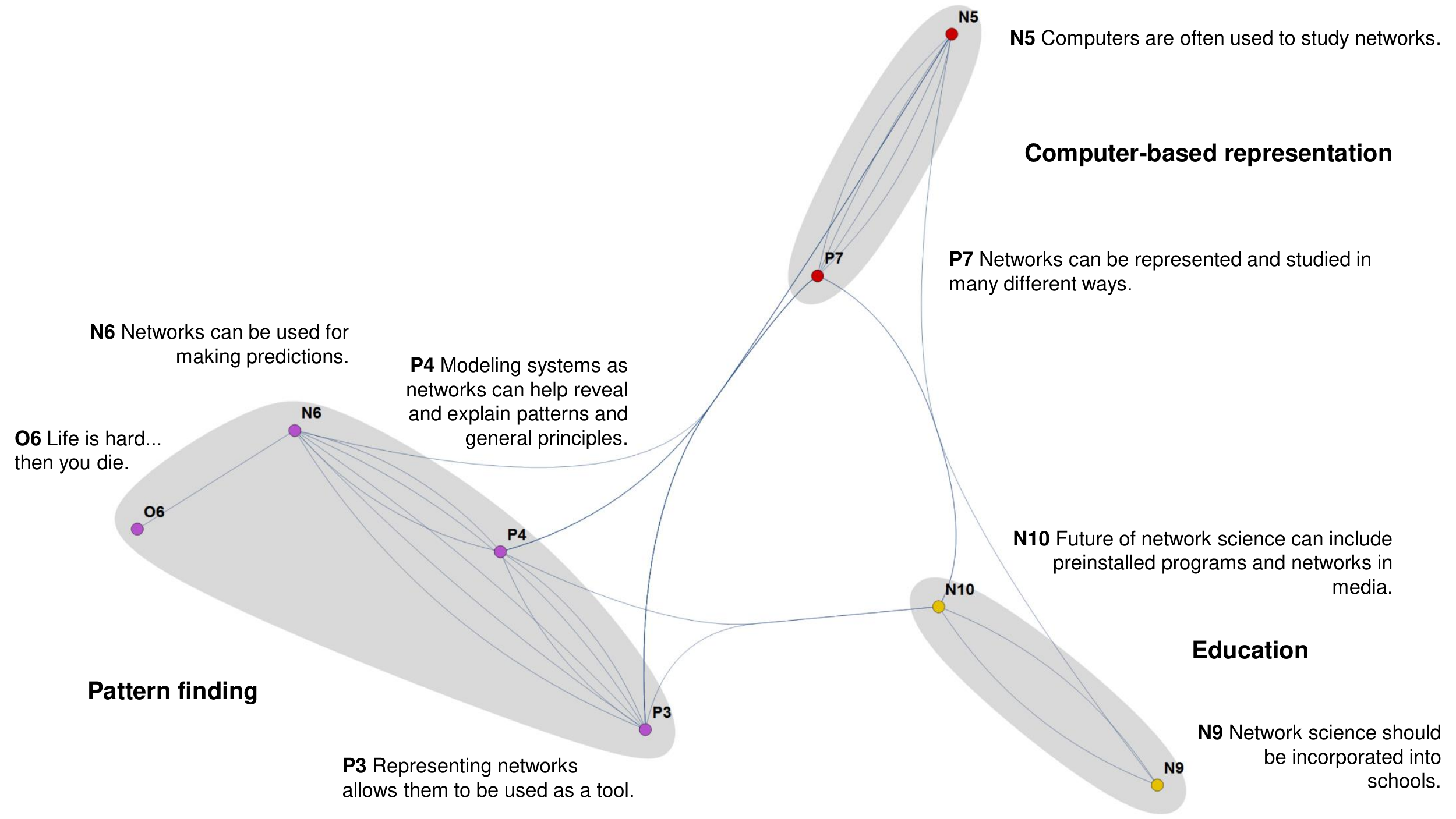}
\caption{Final breakdown of the remaining ``Other'' community in
  Fig.~\ref{fig:comm2}. We find three communities:
  ``Computer-based representation'' (top, new + pro), ``Education'' (bottom
  right, new), and ``Pattern finding'' (bottom left, new + pro + old).
  }
\label{fig:comm3}
\end{sidewaysfigure}

Detecting communities in our final concept multigraph yielded several distinct concept clusters.
These include basic terminology, definitions of networks, ubiquity of networks, and
social applications (see Fig.~\ref{fig:comm1}). Detecting communities in networks consisting of the remaining patchwork communities revealed concept clusters about visualization, computer-based representation, education, and pattern identification
(see Figs.~\ref{fig:comm2} and \ref{fig:comm3}). Three of those clusters (basic terminology, visualization, and education) arose only
in the student groups. This again illustrates
the limitation of brainstorming without active student participation.

The algorithmically-detected concept clusters served as the basis for additional
discussion and refinement of the essential concepts. Through iterative
discussion sessions among the authors, we decided (to reduce potential
redundancy) to include social applications in the cluster about the ubiquity of networks and to combine basic
terminology with definitions of networks. We also decided not to include education explicitly
as a separate concept, because the importance of education was the main
motivation of the Network Literacy initiative. This yielded the following five major concepts:
\begin{enumerate}
\item Ubiquity of networks, including social applications
\item Definitions of networks and basic terminology
\item Pattern identification
\item Visualization
\item Computer-based representation
\end{enumerate}

Through further discussion about the above five concepts, we
recognized that some of the contemporary (and actively-researched) concepts about network
science were missing. Specifically, this includes the power of networks for
interdisciplinary comparison of systems and the interactions between
structures and temporal dynamics of networks. These concepts were present in the list of concepts produced by the professionals, but they were misunderstood and misrepresented in the concept network when the students connected concepts together. This is
possibly because these concepts are among the most difficult ones
about networks for non-professionals to grasp. We decided
to add these concepts to the final list because of their importance.

This yielded a list of seven essential concepts. Incorporating feedback from the network-science community then resulted in the final
version of the seven essential concepts about networks (see Table \ref{tab:final}).

\begin{table}[tbp]
\centering
\caption{Final version of the Network Literacy: Essential Concepts.}
\begin{tabularx}{\textwidth}{lX}
\hline
{\em No.} & {\em Concept}\\
{\bf 1.} & {\bf Networks are everywhere.}\\
{\bf 2.} & {\bf Networks describe how things connect and interact.}\\
{\bf 3.} & {\bf Networks can help reveal patterns.}\\
{\bf 4.} & {\bf Visualizations can help provide an understanding of networks.}\\
{\bf 5.} & {\bf Today's computer technology allows you to study real-world networks.}\\
{\bf 6.} & {\bf Networks help you to compare a wide variety of systems.}\\
{\bf 7.} & {\bf The structure of a network can influence its state and vice versa.}\\
\hline
\end{tabularx}
\label{tab:final}
\end{table}

Naturally, each of the seven concepts subsumes several more detailed
ideas. We thus developed in-depth descriptions of each
concept as a list of ``core ideas''. We also circulated these core ideas
to the network-science community and incorporated their feedback. After numerous iterations of
online and offline edits and reviews of the contents, the current
version of ``Network Literacy: Essential Concepts and Core Ideas'' was
finalized on 4 December 2014, approximately six months after the initial
pre-conference event at NetSci 2014. We include the text of the final version in the Supplemental Material.

\begin{figure}[tbp]
\centering
\fbox{\includegraphics[width=0.4\textwidth]{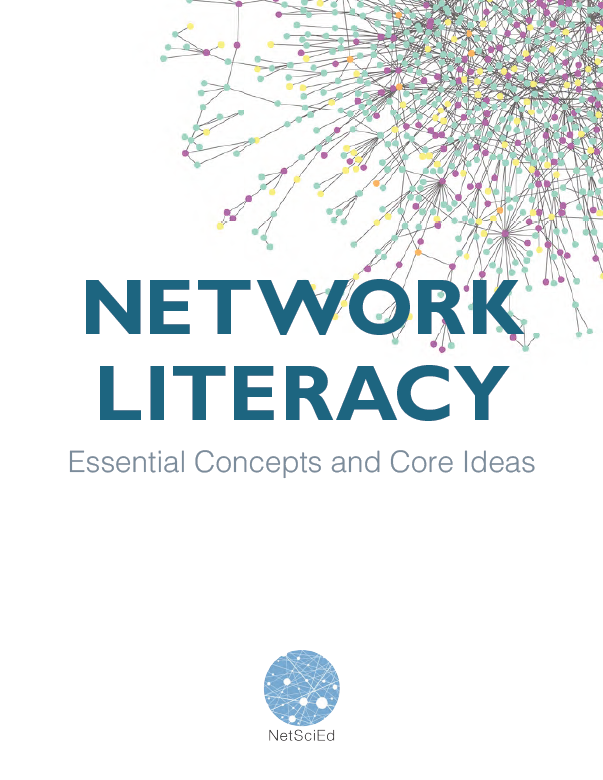}}\\
~\\
~\\
\fbox{\includegraphics[height=0.16\textheight]{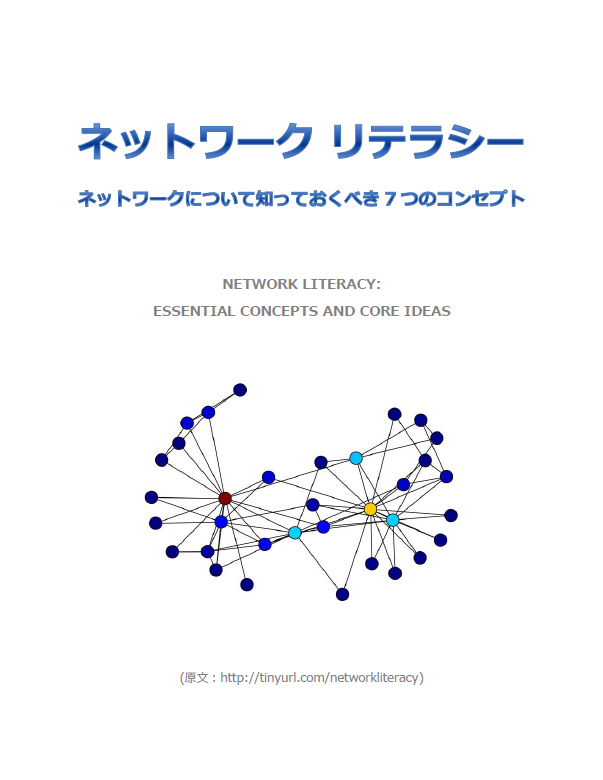}}
\fbox{\includegraphics[height=0.16\textheight]{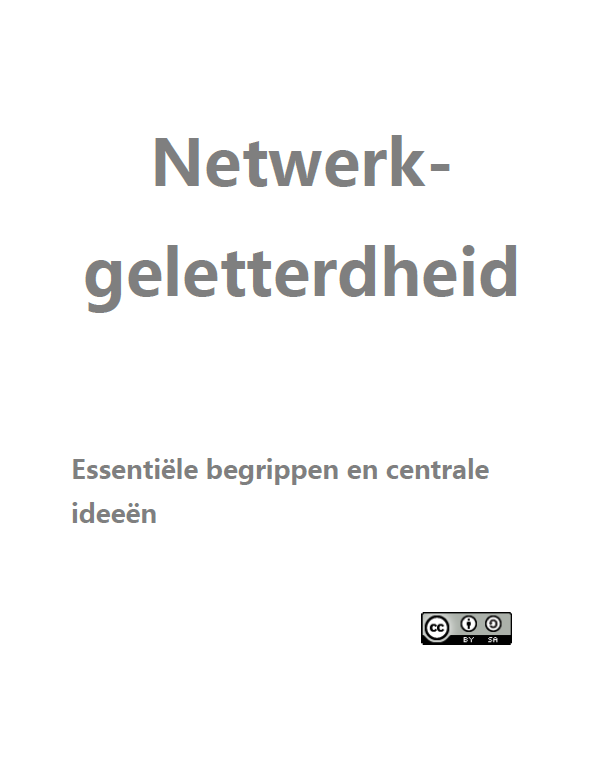}}
\fbox{\includegraphics[height=0.16\textheight]{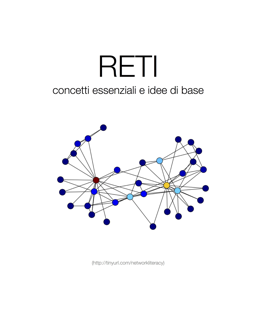}}
\fbox{\includegraphics[height=0.16\textheight]{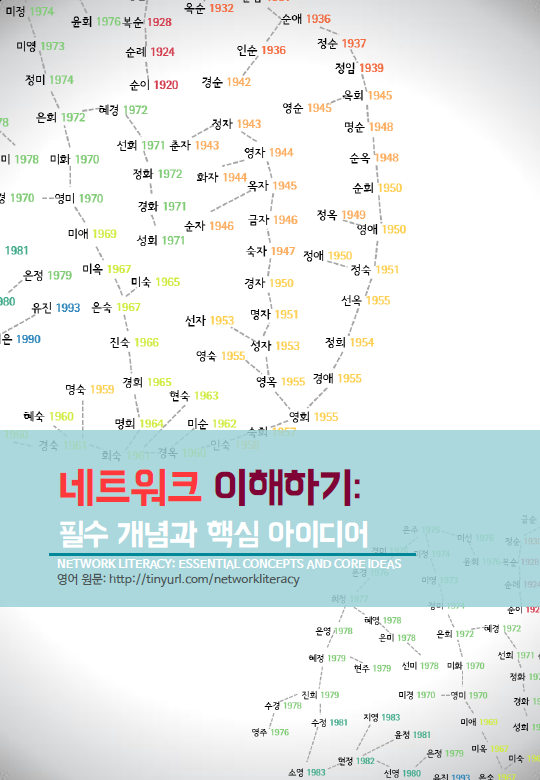}}\\
\fbox{\includegraphics[height=0.16\textheight]{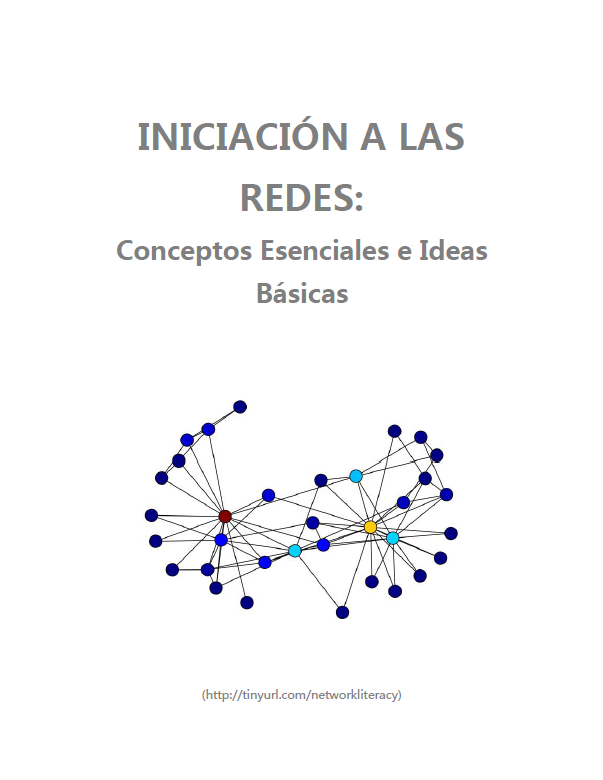}}
\fbox{\includegraphics[height=0.16\textheight]{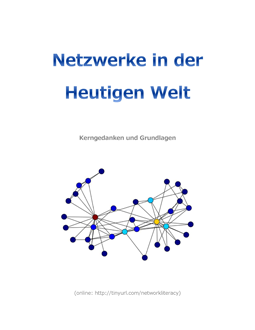}}
\fbox{\includegraphics[height=0.16\textheight]{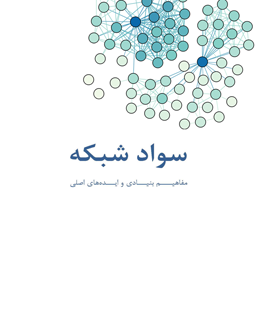}}
\caption{``Network Literacy: Essential Concepts and Core Ideas''
  booklets. Top: Original (English) version. Bottom: Non-English versions
  translated by volunteers. (The Japanese version was produced by Yoshi Fujiwara, Toshihiro Tanizawa, and Hiroki Sayama; the Dutch version was produced by Paul van der Cingel; the Italian version was produced by by Paolo Tieri; the Korean version was produced by Sang Hoon Lee and Mi Jin Lee; the Spanish version was produced by Rosa Benito; the German version was produced by Andreas Joseph and Florian Klimm; and the Persian version was produced by Taha Yasseri.) The graphical design of the original English version was created by Eri Yamamoto; the design of the Korean version was created by Mi Jin Lee; the design of the Japanese version was created by Hiroki Sayama; the design of the Persian version was created by Taha Yasseri based on the English and Japanese versions; and the other four versions adopted the design template of the Japanese one. All of these versions are available at
  \protect\url{http://tinyurl.com/networkliteracy}. Translations into
  other languages are in progress.}
\label{fig:booklets}
\end{figure}

The final text of ``Network Literacy: Essential Concepts and
Core Ideas'' was developed into a professionally designed, printable
booklet for use by teachers, students, and other learners
(see the top panel of Fig.~\ref{fig:booklets}).  It was first published online for
free download at \url{http://tinyurl.com/networkliteracy} on 12 March 2015. Hard copies of the booklet were also disseminated at
CompleNet 2015 on 25 March 2015 and at NetSci 2015 on 1 June 2015. Several very positive responses
followed these initial disseminations. An important development, facilitated through social media and personal contacts, has been the translation of the booklet into non-English languages by volunteers. As of 22 September 2015, seven translated
versions (Dutch, German, Italian, Japanese, Korean, Persian, and Spanish) are available (see the bottom panel of Fig.~\ref{fig:booklets}), and several other translations are in progress.

\section{Conclusions}\label{conclusions}

In this paper, we reported our recent initiative of developing an
educational resource that concisely summarizes essential concepts
about networks in an easily accessible format. The result, ``Network
Literacy: Essential Concepts and Core Ideas'', is the product of a collaborative effort that involved a large number of network-science
researchers, educators, school teachers, and students. Since its initial
release, we have received many very positive responses from across the
globe. We hope that this initiative will help inspire the network-science community
 to make a large societal impact by launching more educational outreach
activities that spread the concepts and ideas about networks into
various formal and informal learning settings---with a particular emphasis on the next
generation, who will live their whole lives in a networked world.

An important lesson from this process is that the deep involvement of students is crucial. The outcomes of the brainstorming sessions
revealed that network-science researchers and educators tended to
consider abstract concepts as ``essential'' but placed less emphasis
on specific examples or research methods (e.g., visualization and
computational tools). However, the students who were exposed to
network science through the NetSci High program frequently mentioned
that concrete examples of networks and the specific research methods that they
learned were among the most essential components of their experience. We believe that such a down-to-earth, hands-on viewpoint
is extremely important when communicating the value of network
science (i.e., ``the study of connectivity'') to the public. Although theoretical abstraction of
systems is crucial for the modeling and analysis of networks, the
general public needs---and wants!---to know {\em what networks are} and
{\em how to use them}. We are confident that our final booklet will play a large role in successfully addressing this issue, and the booklet has benefited greatly from the significant contributions made by high-school
students.

The above successes notwithstanding, our ``Network Literacy'' initiative is limited in
some important respects. First, the booklet that we produced
introduces concepts and ideas only. It does not provide lesson plans
(for teachers) or further study guides (for learners). Developing
lesson plans in particular will require substantially more effort and
resources, as it will need to include details of class instructions,
which needs to sufficiently match local school curricular structures, contents, and
instruction methods (and these are rather heterogeneous) to facilitate adoption in
classrooms. Second, the current list of essential concepts may not
accurately reflect the scientific importance of various aspects of network
science. For example, a major portion of recent discoveries in the study of networks have focused on the dynamical nature of networks and/or on various structural intricacies (e.g., synchronization, diffusion and contagion \cite{portergleeson2014,rom-review2014},
adaptive networks \cite{gross2009adaptive}, temporal networks
\cite{holme2012temporal}, and multiplex and other multilayer networks
\cite{kivela2014multilayer,boccaletti2014structure}), but this topical area was represented
only briefly in concept 7 (and it was not present in
the students' input). How to accessibly convey the richness of complex network
dynamics to the general public remains an important problem to
address. We are currently seeking additional resources to make further
progress on these frontiers of network science and education, and we invite the entire network-science community to
join this challenging yet highly meaningful and societally important endeavor.

\section*{Acknowledgements}

The Network Literacy initiative would not have been possible without
the participation, support, and encouragement of the US Army Research Office, Albert-L\'aszl\'o Barab\'asi, 
Raissa D'Souza, Sarah Schoedinger, H. Eugene Stanley, Craig Strang,
the Network Science Society, NetSci High students and teachers,
the University of California at Berkeley, and all of the members of the
network-science community who have contributed to and supported this
effort. Specifically, the following individuals have made important
contributions to the contents of Network Literacy: Alvar Agusti, Chris
Arney, Robert F. Chen, Arthur Hjorth, Khaldoun Khashanah, Yasamin
Khorramzadeh, Erik Laby, Toshi Tanizawa, Paolo Tieri, Brooke Foucault
Welles, Robin Wilkins, and Eri Yamamoto. 

Translations of ``Network Literacy: Essential Concepts and Core
Ideas'' were made possible thanks to the following volunteers: Paul van der Cingel (Dutch), Yoshi Fujiwara (Japanese),
Toshihiro Tanizawa (Japanese), Paolo Tieri (Italian), Mi Jin Lee (Korean), Sang Hoon Lee
(Korean), Rosa Benito (Spanish), Andreas Joseph (German), Florian Klimm (German), Taha Yasseri (Persian), Ralucca Gera (Romanian, in progress), and Peter Polner (Hungarian, in progress). 

The authors also thank Katherine Culp, Amanda Jaksha and Matty Lau for their valuable feedback on Fig.~\ref{fig:workflow}.

The material in this paper is based on work supported in part by the US National Science Foundation under Grant No. 1027752 and 1139478.

\bibliographystyle{plain}
\bibliography{network-literacy}

\newpage
\appendix

\section*{Supplemental Material: Final Text of Network Literacy: Essential Concepts and Core Ideas}

\subsection*{1: Networks are everywhere.}
\begin{itemize}
\item The concept of networks is broad and general, and it describes
  how things are connected to each other. Networks are present in
  every aspect of life.
\item There are networks that form the technical infrastructure of our
  society---e.g., communication systems, semantic systems, the
  Internet, electrical grids, the water supply, etc.
\item There are networks of people---e.g., families and friends,
  e-mail/text exchanges, Facebook/Twitter/Instagram, professional
  groups, etc.
\item There are economic networks---e.g., networks of products,
  financial transactions, corporate partnerships, international
  trades, etc.
\item There are biological and ecological networks---e.g., food webs,
  gene/protein interactions, neuronal networks, pathways of disease
  spreading, etc.
\item There are cultural networks---e.g., language/literature/art
  connected by their similarities, historical events linked by causal
  chains, religions connected by their shared roots, people connected
  to events, etc.
\item Networks can exist at various spatial and/or temporal scales.
\end{itemize}

\subsection*{2: Networks describe how things connect and interact.}
\begin{itemize}
\item There is a subfield of mathematics that applies to networks. It
  is called {\em graph theory}. Many networks can be represented
  mathematically as {\em graphs}.
\item Connections are called {\em links, edges, or ties}. The entities
  that are connected to each other are called {\em nodes, vertices, or
    actors}.
\item Connections can be undirected ({\em symmetric}) or directed
  ({\em asymmetric}). They can also indicate ties of different
  strengths, and can indicate either positive or negative
  relationships.
\item The number of connections of a node is called the
  {\em degree} of that node.
\item Many networks have more than one type of connection---e.g.,
  offline friendships and Facebook connections, different modes of
  transportation, etc.
\item A sequence of edges that leads from one node, through other
  nodes, to another node is called a {\em path}.
\item A group of nodes within which a path exists from any one entity
  to any other entity is called a {\em connected component}. Some
  networks have multiple connected components that are isolated from
  each other.
\item Some networks are studied using mathematical structures that are
  more complicated than graphs.
\end{itemize}

\subsection*{3: Networks can help reveal patterns.}
\begin{itemize}
\item You can represent something as a network by describing its parts
  and how they are connected to each other. Such a network
  representation is a very powerful way to study a system's
  properties.
\item The properties in a network that you can study include:
\begin{itemize}
\item how the degrees are distributed across nodes
\item which parts or connections are most important
\item strengths and/or weaknesses of the network
\item if there is any sub-structure or hierarchy
\item how many steps, on average, are needed to move from one node to
  another in the network
\end{itemize}
\item In some networks, you can find a small number of nodes that have
  much larger degrees than others. They are often called {\em hubs}.
\item In some networks, you can find a group of nodes that are better
  connected to each other than chance would dictate. They are
  sometimes called {\em clusters} or {\em communities}. Some of them
  can occupy a central, or {\em core}, part of a network.
\item Using these findings, you can sometimes infer how a network was
  formed and/or make predictions about dynamical processes on the
  network or about its future structure.
\end{itemize}

\subsection*{4: Visualizations can help provide an understanding of networks.}
\begin{itemize}
\item Networks can be visualized in many different ways.
\item You can draw a diagram of a network by connecting nodes to each
  other using edges.
\item There are a variety of tools available for visualizing networks.
\item Visualization of a network often helps to understand it and
  communicate ideas about connectivity in an intuitive, non-technical
  way.
\item Creative information design plays a very important role in
  making an effective visualization.
\item It is important to be careful when interpreting and evaluating
  visualizations, because they typically do not tell the whole story
  about networks.
\end{itemize}

\subsection*{5: Today's computer technology allows you to study real-world networks.}
\begin{itemize}
\item Computer technology has dramatically enhanced the ability to
  study networks, and this is especially important for large ones with
  rich structure.
\item There are many free software tools available for network
  visualization and analysis.
\item Using personal computers, everyone (not just scientists) can
  construct, visualize, and analyze networks.
\item Through the Internet, everyone has access to many interesting
  network data sets.
\item Computers allow you to simulate hypothetical or virtual
  networks, as well as to simulate dynamical processes on both real
  and hypothetical networks.
\item Learning computer literacy skills opens the door to myriad
  possibilities for a career. These include scientist, data analyst,
  software engineer, educator, web developer, media creator, and many
  others.
\end{itemize}

\subsection*{6: Networks help you to compare a wide variety of systems.}
\begin{itemize}
\item Various kinds of systems, once represented as networks, can be
  compared to examine their similarities and differences.
\item Certain network properties commonly appear in many seemingly
  unrelated systems. This implies that there exist some general
  principles about connectivity that apply to multiple domains.
\item Other network properties are different in different
  systems. These properties can help to classify networks in different
  families and to gain insight into why they are different.
\item Science is typically conducted in separate areas of research
  called disciplines. Networks can help to cross disciplinary
  boundaries to achieve a holistic and more complete understanding of
  the world.
\item Networks can assist in the transfer of knowledge across
  different areas of study.
\end{itemize}

\subsection*{7: The structure of a network can influence its state and vice versa.}
\begin{itemize}
\item Network structure indicates how parts are connected in a
  network.
\item Network state indicates the properties of a network's nodes and
  edges.
\item Network structure and state can each change over time.
\item The time scales on which network structure and state co-evolve
  can be either similar or different.
\item Network structure can influence changes of network
  state. Examples include the spread of diseases, behaviors, or memes
  in a social network, and traffic patterns on the road network in a
  city.
\item Network state can influence changes of network
  structure. Examples include the creation of new ``following'' edges
  in social media and the construction of new roads to address traffic
  jams.
\end{itemize}

\end{document}